# Performance Estimation of 2*3 MIMO-MC-CDMA in Rayleigh Fading Channel

Mr. Atul Singh Kushwah & Mr. Sachin Manglasheril

*Asst. Professor & Electronics & Communication Department & Indore Institute of Science & Technology-II, Indore (M.P), India*

*Abstract -* **In this paper we analyze the performance of 2*3 MIMO-MC-CDMA system in MATLAB which greatly reduces BER by increasing the efficiency of system. MIMO and MC-CDMA system arrangement is used to decrease bit error rate and also figure a new system called MC-CDMA which is multiple user and multiple access system used to enhance the performance of the system. MC-CDMA is a narrowband flat fading in character which changes frequency selective to several narrowband flat fading multiple parallel sub-carriers to enhance the effectiveness of the system. Now this MC-CDMA further improved by grouping by 2*3 MIMO system which make use of ZF decoder at the receiver to reduce BER in which half-rate convolutional encoded Alamouti's STBC block code is intended for channel encoding as transmit diversity for MIMO with multiple transmit and receive antennas. Main improvement of using MIMO-MC-CDMA is for reducing complication of system and also for dropping BER and finally increasing gain of the system. Then we estimate the system in different modulation techniques like, 64-QAM, 8-PSK, 16-QAM, QPSK, 32-QAM and 8-QAM using MATLAB in Rayleigh fading channel.**

## I. INTRODUCTION

Due to modern necessity of technology for high data rate and low probability of error in this paper we join systems like OFDM, CDMA and MIMO, which forms advanced technique for declining bit error rate. MC-CDMA is produced by combination of OFDM and CDMA which is multiple access and multi-carrier systems [11]. MC-CDMA (Multiple Carrier- Code Division for Multiple Access) is narrowband flat fading channel. The MC-CDMA increases the performance of wireless communication system through high data rate and low probability of error.

In this paper we join MC-CDMA with MIMO to additional increase in performance. We use 2*3 MIMO antenna diversity system which is also called is multiple antenna scheme in which two transmit and three receive antennas are used and for detection of symbols at receive diversity and transmit diversity we make use of half-rate convolutionally encoded Alamouti STBC block code which is used for synchronization of system to reduce Inter Symbol Interference. For orthogonality recognition of signal Zero-Forcing(ZF) detection method is used. Finally we got MIMO-MC-CDMA [4] by combination of above explained system and this combination is prepared by using MATLAB then performance analysis is done by using various modulation techniques like 16-QAM, 8-PSK, QPSK, 32-QAM, 8-QAM and 64-QAM in Rayleigh fading channel.

## II. MIMO OVERVIEW

MIMO refers to Multiple Input and Multiple Output in which multiple antennas are used in MIMO [2] system at both transmitter and receiver and both diversity are functional to reduce fading resulting from signal variations through the wireless channel. The system offer diversity gains based on the degree at which the multiple data replicas are to be faded independently which represents the variation in SNR at the output of diversity combiner as evaluated to single branch diversity by the face of certain probability level. MIMO system include N number of transmit antenna that is equal to two, and of M number of receive antenna elements that is equal to three was modeled, according to that diversity order of 6 could be achieved. Merge the numerous versions of the signals produced by special diversity schemes is advantageous for optimizing the performance. This paper utilizes zero forcing (ZF) technique as a decoder to merge M received signals to reverberate on the most likely transmitted signal. The measure of the received SNRs from M different paths is the proficient received SNR of the system by means of diversity order M. The main requirements of receiver is to demodulate all M received signal in terms of ZF for basis through M independent signals at the received antennas.

### III. MULTI-CARRIER CDMA

MC-CDMA [7] is the arrangement of OFDM and CDMA, consequently better frequency diversity and improved data rates. In MC-CDMA, each symbol is spreaded by code chips and transmitted by some subcarriers. It is not essential that the number of carriers to be equivalent to the code length consequently providing a degree of flexibility in our design. In MC-DS-CDMA information is spreaded in time domain relatively frequency domain. In MC-CDMA particular data symbol is transmitted through independent subcarriers. The significant advantage of MC-CDMA is improve the bandwidth efficiency through multiple access is possible by appropriate system design by orthogonal codes.

#### A. Need of MC-CDMA

MC-CDMA employ the benefits of both OFDM and CDMA and forms an efficient transmission system by the spreading of input data symbols through spreading codes in frequency domain. The amount of narrowband orthogonal subcarriers through symbol period longer than the delay spread and each subcarrier are affected by similar deep fades by the channel at the same time cause enhance the performance. Adding to the number of path will enhance the performance of system this is enhanced mainly by two reasons firstly due to diversity, then, it decline due to the increase in the interference from various paths at all users. Generally, there are proficient number of paths that based on the system to be used and the number of users. Interference will be increased with the number of users through all the paths. So, the best possible number of paths decreased.

#### B. MCCDMA System Model

MC-CDMA [3,5,6] transmitter and OFDM transmitter have slight difference. In OFDM many symbols are transmitted by subcarriers however in MC-CDMA like symbol is transmitted by different subcarriers. The block diagram of MC-CDMA is revealed in fig.1. The input information symbols are transformed into parallel stream of information. The OFDM system coupled by means of the CDMA system transfer the symbols to time domain by IFFT and allot subcarrier used for all symbol. After that the subcarriers were multiplexed to form as a serial stream of information. Prior to the transmission the serial stream that is transformed into blocks and each block is separated by a guard frame. The guard frame might be zero symbols or called as padding symbols. In OFDM the CP is used as guard symbols that have a variety of compensation to remove ISI and ICI caused by multipath fading. So the cyclic prefix length such that it is larger than the delay spread of channel. In MC-CDMA transmission, which is frequency non selective fading over all sub carrier. So, if the individual symbol rate is more enough to convert it into frequency selective fading, the input data will be serial to parallel (S/P) converted into parallel data sequences and each S/P output is then multiplied by spreading code of unusual length. For enhancing the performance of the system, an appropriate approach is used for channel analysis, to make use of dedicated pilot symbols that are inserted occasionally in the transmission which also called as block-type pilot channel estimation.

In MCCDMA receiver pattern planned for the jth user as shown in Fig.1. The received signal is initially down converted. Then, removing the cyclic prefix and the residual samples are transformed serial to parallel to achieve the subcarriers components. The subcarriers are initially demodulated by a FFT and after that multiplied by the gain to join the received signal energy scattered in the frequency domain.

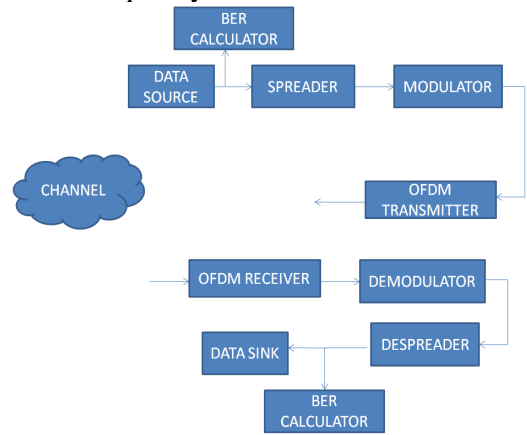

Fig.1 : Block Diagram of MC-CDMA

## IV. MIMO-MC-CDMA SYSTEM MODEL

Communication system representation of MIMO-MC-CDMA [1] is depicted in fig.2.

In MIMO-MC-CDMA communication system we assume that transmitter transmit random string of data to the receiver so we employ random PN sequence generator by MATLAB. Now spreading of that sequence is done by using PN sequence generator for indicating random data input through users. After that modulator is employed for various modulation schemes are employed like *32-QAM, 8-QAM, 8-PSK, 16-QAM, QPSK and 64-QAM* this is represented by modulator block. MC-CDMA system is already detailed in section III by Multi-Carrier Code Division Multiple Access. Now MIMO by means of half rate convolutionally encoded Alamouti's STBC block code is employed which is explained in section II as Multiple Input Multiple Output (MIMO). The combination of MIMO and MC-CDMA outline the improved system model called MIMO-MC-CDMA as shown in fig.2. After that signal send through Rayleigh Fading Channel [7]. Then receiver receive the signal in reverse approach with ZF decoder for the renewal of transmitted information at the receiver and BER computation is done for estimate the system performance. In MIMO system two transmit and three receive antennas are utilized. In this paper we are transmitting the information bits which are random in character or information dependent on user then the data is passed by the spreader using PN sequence generator that forms 8 bits for every input bit then resulting bits are created after the spreading of encoded information sequence. After that these spreading sequence are passed to modulator in which their modulation depends upon the different modulation scheme to be used. Then this modulated data is reframed to parallel data for OFDM then IFFT is employed to change frequency selective carriers into parallel narrowband flat-fading carriers that are orthogonal in character, then this is again transformed into parallel to serial after that CP cyclic prefix is added to eradicate ISI which complete the process of OFDM transmission scheme, then serial data passed by MIMO with Alamouti STBC code designed for 2 transmit and 3 receive diversity antenna systems in which 3*2 channel matrix is produced by using MIMO diversity, and also the ZF detection method is used at the receiver to sense orthogonality after that reverse process is done for receiving input bits at receiver.

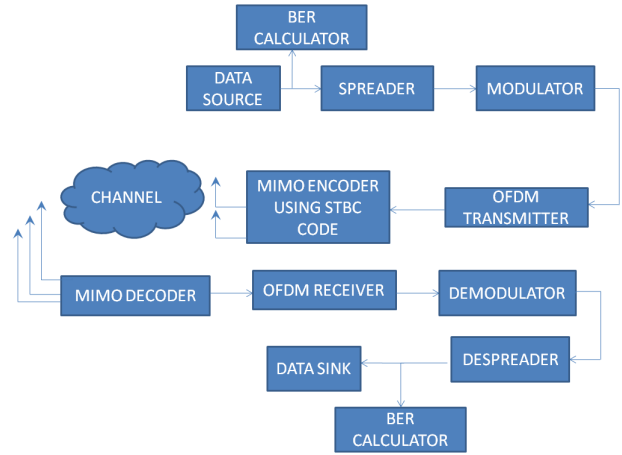

Fig.2. Communication System Model OF 2*3 MIMO-MC-CDMA

## V. SIMULATION RESULTS AND DISCUSSION

Table 1 shows input parameters for MIMO-MC-CDMA [7,8,9,10] using different modulation schemes.

Fig.3 shows the relative analysis of different modulation techniques in MIMO-MC-CDMA.

Table 2 depicts the performance estimation of different modulation techniques in terms of gain and BER.

TABLE I

**SIMULATED MODEL PARAMETERS**

| Channel Encoder | ½ rate convolution encoder Alamouti STBC |
|---|---|
| Signal to Noise Ratio | 0dB to 20 dB |
| CP Length | 1280 |
| OFDM Sub-carriers | 6400 |
| No. of transmitting and receiving antennas | 2*3 |
| Modulation Schemes | QPSK, 8-PSK, 8-QAM, 16-QAM, 32-QAM and 64 QAM |
| Channel | Rayleigh Fading Channel |
| Signal detection scheme | Zero forcing |

From table 2 and Fig.3 we can observe that QPSK shows high gain of 25.91 dB with very low BER as compared to different modulation schemes at 0dB SNR. This is employed by using MIMO-MC-CDMA system by which as result error probability in QPSK is nearly zero which shows very low probability of error in system.

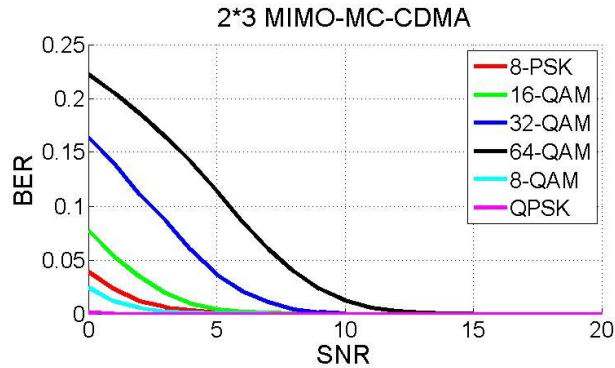

Fig.3. Performance estimation of 2*3 MIMO-MC-CDMA in different modulation scheme.

TABLE II

**PERFORMANCE ANALYSIS AT 0DB SNR WITH RESPECT TO 64-QAM MODULATION TECHNIQUE AS SHOWN IN FIG.3**

| Modulation | BER at 0dB SNR | Gain w.r.t 64-QAM |
|---|---|---|
| QPSK | 0.00057 | 25.91 dB |
| 8-QAM | 0.0241 | 9.64 dB |
| 8-PSK | 0.03897 | 7.56 dB |
| 16-QAM | 0.0774 | 4.57 dB |
| 32-QAM | 0.164 | 1.31 dB |
| 64-QAM | 0.2221 | 0dB |

VI. CONCLUSION

Fig.3 represents the relative estimation of MIMO-MC-CDMA in various modulation schemes. Table 2 shows the relative analysis used for various modulation schemes which shows that as modulation order is improved results increase in BER. This paper aims to reduced bit error rate which is done by QPSK modulation scheme and resultant gain of 25.91 dB with respect to 64-QAM modulation technique which prove that the gain of QPSK is highest as compared to other modulation technique with extremely low probability of error because errors are nearly finished at 0dB in QPSK modulation scheme. For 3G and 4G wireless communication system 64-QAM modulation scheme is favoured which hold BER upto 10dB, i.e. errors are remain in 64-QAM upto 13dB SNR which is improved by using MIMO-MC-CDMA.